\documentclass{aa}
\usepackage{psfig}

\def\object{{}}
\begin{document}
\title{$^{13}$CO at the centre of M\,82}
\thesaurus{03 (09.02.1; 11.09.1 M82; 11.09.4; 11.11.1; 11.19.3; 13.19.1)}  %
\date{Received August 14, 1997, accepted September 1, 1998}  
\author{N. Neininger\inst{1,2,3}
 \and M. Gu\'elin\inst{3}
 \and U. Klein\inst{1}
 \and S. Garc\'{\i}a-Burillo\inst{4}
 \and R. Wielebinski\inst{2} }
%
%
\institute{ 
           Radioastronomisches Institut der Universit\"at Bonn,
           Auf dem H\"ugel 71, D--53121 Bonn, Germany  
     \and  Max--Planck--Institut f\"ur Radioastronomie,
           Auf dem H\"ugel 69, D--53121 Bonn, Germany
     \and  Institut de Radioastronomie Millim\'etrique,
           300 rue de la Piscine, F--38406 St.\ Martin d'H\'eres, France 
     \and  Observatorio Astron\'omico Nacional, 
           Apartado 1143, E--28800 Alcal\'a de Henares, Spain
          }
\maketitle
%
%
\begin{abstract}
Using the IRAM interferometer, we have observed the nearby starburst 
galaxy M\,82 with a $4\farcs2$ resolution ($\simeq 70$ pc) in the 
1$\rightarrow$0 line of $^{13}$CO and in the $\lambda$ 2.6-mm continuum.

The spatial distribution of the $^{13}$CO line shows the same gross 
features as the $^{12}$CO(1$\rightarrow$0) map of Shen \& Lo (\cite{shen}),
namely two lobes and a compact central source, though with 
different relative intensities. The lobes are more 
conspicuous and the central source is fainter in $^{13}$CO than in 
$^{12}$CO. 

The velocity field observed around the nucleus shows a very steep 
gradient (140 km\,s$^{-1}$ over 75\,pc), which is very probably 
caused by the stellar bar visible in the near infrared. 
The dynamical centre coincides with the IR peak and is shifted 
$6''$ north-east of the compact $^{13}$CO source. 
The two CO lobes appear to be associated with the ends of the bar 
and not with a molecular ring, as usually assumed. They are probably 
shaped by the strong UV radiation from the central region. $^{13}$CO
must be more photodissociated than the self-shielded $^{12}$CO
molecules in the central $\sim$250~pc region, which may explain 
the relative weakness of the $^{13}$CO central source.

A 130 pc-wide bubble of molecular gas has been identified, which 
happens to host the most luminous compact radio source in M\,82.  
It lies 120 pc west of the IR peak between the central source 
and the western lobe and seems characterized by warmer gas, strong UV
radio free-free radiation, and an enhanced cosmic ray production 
rate.  

  \keywords{ISM: bubbles -- Galaxies individual: M\,82 -- Galaxies: ISM 
   -- Galaxies: kinematics and dynamics -- Galaxies: starburst -- 
   Radio lines: galaxies}  %
\end{abstract}

\section{Introduction} \label{sec:intro}

The central kpc region of \object{M\,82}, the archetype of starburst 
galaxies, provides a unique laboratory to study 
intense star formation from an assembly of giant molecular clouds. 
The starburst, possibly triggered by a tidal interaction with \object{M\,81},
manifests itself through a high far-infrared (far IR) luminosity and 
a high density of supernova remnants (SNR) (see e.g.\ 
Telesco \& Harper \cite{telesco8}; Kronberg et al.\ \cite{kronberg}). 
The presence of copious molecular gas is obvious from the many 
single-dish studies of the CO mm lines (Nakai et al.\ 
\cite{nakai}; Olofsson \& Rydbeck \cite{olofsson}; Loiseau et al.\ 
\cite{loiseau8}, \cite{loiseau9}). A large fraction of this gas is 
concentrated in massive hot and dense clouds, which give rise to conspicuous
emission in dozens of high-excitation molecular lines, such as the
high-J lines of CO and the lines of CS and HCN (Henkel \&  
Bally \cite{henkel}, Wild et al.\ \cite{wild}). These clouds, 
according to multi-transition analysis, have temperatures of $\sim $40 K and 
densities of few 10$^4$ cm$^{-3}$ (Wild et al.\ \cite{wild}; 
G\"usten et al.\ \cite{guesten}). Interferometric observations
show that they have a patchy distribution (Lo et al.\ \cite{lo}; 
Brouillet \&  Schilke \cite{brouillet}; Shen \&  Lo \cite{shen}).

The molecular gas at the centre of \object{M\,82} is 
usually thought to be concentrated in a rotating circumnuclear torus 
(e.g.\ Nakai et al.\ \cite{nakai}, Shen \&  Lo \cite{shen}).
The presence of a stellar bar (Telesco \& Gezari 1992) may 
explain how gas is driven inwards in order to fuel the starburst. 
The main activity is however already subsiding according to Rieke et 
al.\ (\cite{rieke}, see also Shen \&  Lo \cite{shen}); this may seem 
surprising given the large amount of molecular gas in the ``ring''.

Multi-transition analysis show that the $^{12}$CO mm lines are 
optically thick in \object{M\,82} (Wild et al.\ \cite{wild}),
which makes it hazardous to study the gas distribution and kinematics
of this edge-on galaxy from this isotopomer alone. The H{\sc i} line,
moreover, gives little insight into the kinematics of the central
regions as its profile shows a mixture of emission and absorption
components. 

\object{M\,82} has been mapped in the optically thin $^{13}$CO and 
C$^{18}$O lines (Loiseau et al.\ \cite{loiseau8}, 
\cite{loiseau9}; Wild et al.\ \cite{wild}), but with spatial resolutions 
of at best 150~pc. Both the $^{13}$CO brightness distribution
and velocity field appeared markedly different from those observed in 
$^{12}$CO, an unusual result at such low resolutions.
Single-dish studies of compact and weak extragalactic sources, however,
are hampered by insufficient resolution and pointing errors. An 
interferometric study in the optically thin 
$^{13}$CO line was therefore mandatory. Here we present such a 
study carried out with the IRAM Plateau de Bure 
interferometer. In Sect.~\ref{sec:obse} we describe 
the observations and data analysis. Sect.~\ref{sec:result} presents 
the distribution and kinematics of the $^{13}$CO emission. A prominent 
molecular arc-structure is described and dicussed in 
Sect.~\ref{sec:arc}. Sect.~\ref{sec:discuss} analyzes the 
observations in the frames of the circumnuclear torus and bar scenarios. 
Finally, we give our conclusions in Sect.~\ref{sec:sum}.

\section{Observations and data reduction} \label{sec:obse}

The central region of \object{M\,82} was observed in the 
1$\rightarrow$0 line of $^{13}$CO ($\nu_{\rm obs}$ = 110.116824~GHz) 
with the IRAM interferometer at Plateau de Bure. The surveyed area 
is centred on the $2.2\mu$m nucleus (at $\alpha =  09^h51^m43\fs4, 
\delta = 69\degr55''00''$ B1950.0), Joy et al.\ (\cite{joy}); see also 
Lester et al.\ (\cite{lester}) and consists of three overlapping 
fields corresponding to three pointings of the $45''$ FWHM primary beam, 
shifted, respectively, by ($\alpha ,\delta$)=($-20'', -9''$), ($0'', 0''$), 
and ($+20'', +9''$) with respect to the nucleus. These fields were 
successively observed for 6 minutes each, the observing sequence being 
completed by a 4 min.\
integration on a phase calibrator (0716+714). We used the `compact' CD 
antenna configuration, with a longest baseline of 176~m and a shortest 
one of 24~m. The six units of the correlator were combined to cover a 
1500~km~s$^{-1}$-wide band, with a velocity resolution of 
6.4\,km\,s$^{-1}$ and a 40~km~s$^{-1}$-wide band centred on the source 
systemic velocity, $V_{sys}(LSR)=225$ kms$^{-1}$, with a 0.4~km~s$^{-1}$ 
resolution. The continuum level was derived in the lower sideband of the 
receiver (110\,GHz) from the outer channels ($|V(LSR)-V_{sys}(LSR)|\geq 200$ 
kms$^{-1}$) of the broad band and in the upper sideband (113\,GHz) from 
the entire 1500~km~s$^{-1}$-wide band.

The observations were carried out in August 1994 (4 antennas, D 
configuration) and October 1994 (3 antennas, C1+C2 configurations); 
the D configuration observations were repeated in April 1995 (4 
antennas). For the calibration of the data we used 
0923+392, 3C84, 3C345 and 3C454.3 as primary amplitude and RF bandpass 
calibrators. In the data reduction process, the transfer function was 
tapered with a 90-m (FWHP) Gaussian to yield a circular synthesized 
beam of size $4\farcs2$, which at the adopted distance of M~82 (3.25 Mpc). 
corresponds to 66 pc. The 3 fields were combined as a mosaic and 
subsequently cleaned using the MAPPING procedure of the GILDAS 
package; this yields a roughly constant sensitivity along the line 
joining the centres of the 3 fields.

\section{Results} \label{sec:result}

\subsection{Continuum emission at 110~GHz} \label{sec:conti}

\begin{figure*}    
\vbox{ 
\psfig{file=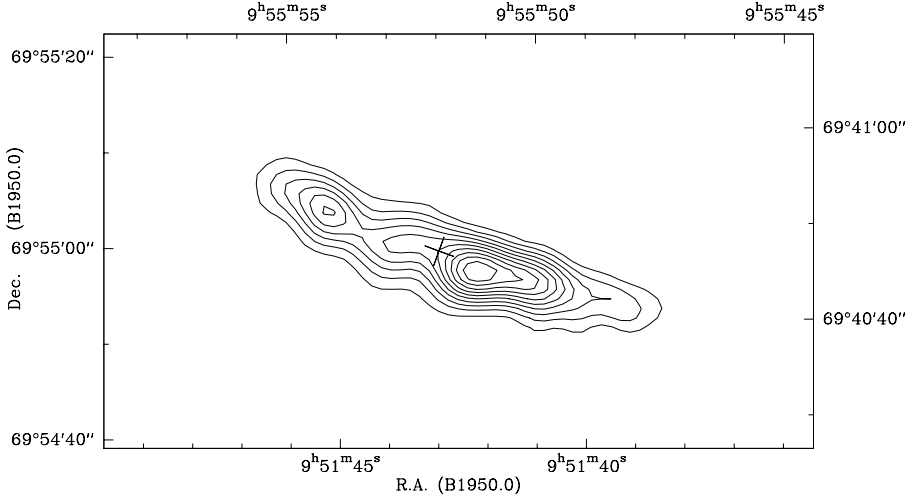,width=12cm,angle=270}\vspace{-4.0cm}}
\hfill\parbox[b]{5.5cm}{
\caption
{Map of the continuum emission of M\,82 at 2.7~mm. The 
contour interval is 5~mJy/beam (0.03~K). The cross marks the 
position of the $\lambda 2.2~\mu$m nucleus derived by Joy et al.\ 
(\cite{joy}). Note that in this figure the J2000.0 coordinates are given 
for comparison at the upper and right sides of the box. The field of view 
is identical to those of Figs.\  2, 3 and 4. } 
\label{fig:cont}
}
\end{figure*}

Fig.~\ref{fig:cont} shows the continuum map constructed from the 
line emission-free channels of the upper (113\,GHz) and lower 
(110\,GHz) receiver sidebands. The average wavelength is close to 
2.7~mm. The rms noise in this map is 2.5~mJy/beam (14~mK), its 
resolution $4\farcs2$. Comparison with the   
continuum maps of Carlstrom \& Kronberg (\cite{carlstrom}), Brouillet 
\& Schilke (\cite{brouillet}) and Seaquist et al.\ (\cite{seaquist9}) 
shows a tight correspondence between all the four interferometric 
maps in the range 90--110 GHz. 

The mm continuum emission in M~82 is a mixture of thermal free-free,
nonthermal synchrotron, and thermal dust emissions (Klein et al.\ 1988). 
Free-free emission is expected to dominate largely at $\lambda$~3~mm, the
relativistic electrons having lost too much energy to radiate
significantly at this short wavelength; therefore, one expects a
nearly flat spectrum (see Carlstrom \& Kronberg \cite{carlstrom}). 
Integrating the flux density over our 
110\,GHz map, we derive a value of $0.55 \pm 0.05$~Jy, in good 
agreement with the values measured at lower frequencies by other groups 
(Jura et al.\ \cite{jura}; Carlstrom \& Kronberg \cite{carlstrom}; 
Brouillet \& Schilke \cite{brouillet}; Seaquist et al.\ \cite{seaquist9}).
This confirms the thermal free-free nature of the continuum radiation.  

\subsection{$^{13}$CO line emission} \label{sec:line13co}

\begin{figure*}        
\psfig{file=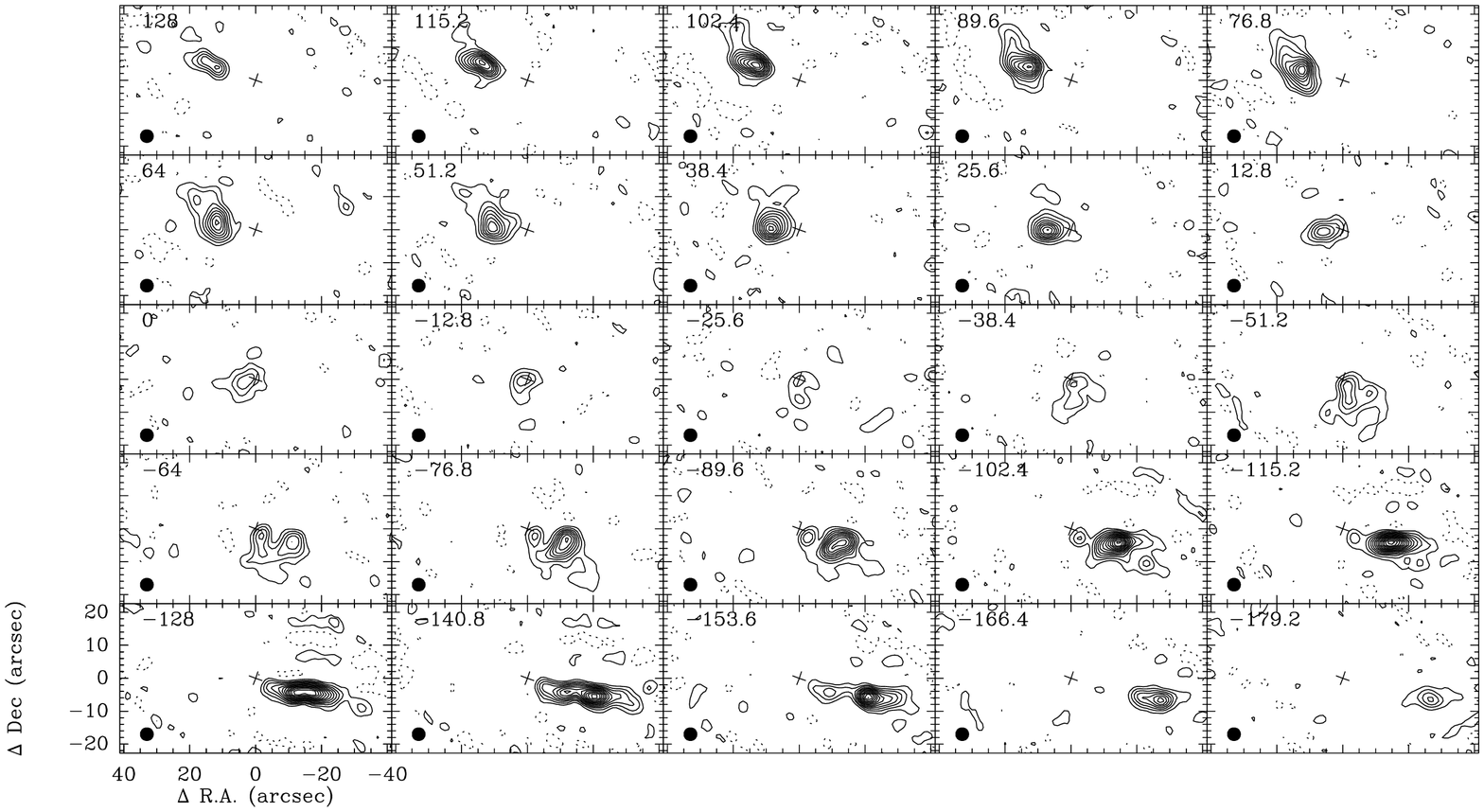,width=18cm,angle=270} 
\caption
{$^{13}$CO(1$\rightarrow$0) velocity-channel maps. The offsets in R.A. and 
Dec are relative to the $\lambda2~\mu$m nucleus which is denoted by a cross 
(R.A. $09^h51^m43\fs4$, Dec $69\degr55'00\farcs0$ B1950.0, Joy et al.\ 
\cite{joy}); E is left, N up. Each velocity-channel is 12.8 km\,s$^{-1}$ 
wide; its centre velocity is indicated in the top left corner in units of 
km\,s$^{-1}$. Velocity 0 corresponds to the galaxy systemic velocity, 
$V_{sys}(LSR)= +225$~km\,s$^{-1}$. The intensities are corrected for 
attenuation by the primary beam. The first contour level and the contour 
interval are 20~mJy/beam (0.1~K). The black dot in the bottom left corner 
of each map represents the half-power synthesized beam.} 
\label{fig:chann13}
\end{figure*}

In Fig.~\ref{fig:chann13} we show the CLEANed velocity-channel maps\footnote 
{Unless specified otherwise, all velocities quoted in this paper are relative 
to the systemic velocity  $V_{sys}(LSR)= +225$~km\,s$^{-1}$.} between 
$+128$~km\,s$^{-1}$ and $-179$~km\,s$^{-1}$. The velocity resolution has 
been degraded to 12.8\,km\,s$^{-1}$ for the sake of clarity. These 
channel maps cover the velocity range within which strong $^{13}$CO 
emission was found. Significant emission was actually detected over a 
larger velocity range (+140~km\,s$^{-1}$ to -190 ~km\,s$^{-1}$). 
Positive velocity-channels are dominated by an emission ``lobe'', 
centred $\simeq$ 15$''$E of the nucleus. This lobe stays at about the 
same place from +130 to $+60$~km~s$^{-1}$, while its position angle turns 
from $\sim70\degr$ to $\sim45\degr$. Near +100~km~s$^{-1}$, a spur is seen 
to emerge to the N of the lobe. Between $-64$~km\,s$^{-1}$ and 
$-90$~km\,s$^{-1}$ an arc-like structure is clearly visible. The negative
velocity-channels are however dominated by a second emission lobe, W of 
the nucleus. It extends further out and to larger velocities than the 
eastern lobe; its orientation is almost E-W. 
Emission is also detected within 5$''$ of the nucleus. It extends from 
$+26$~km\,s$^{-1}$ to $-154$~km\,s$^{-1}$, a remarkably large velocity range.

\begin{figure*}    
\vbox{ 
\psfig{file=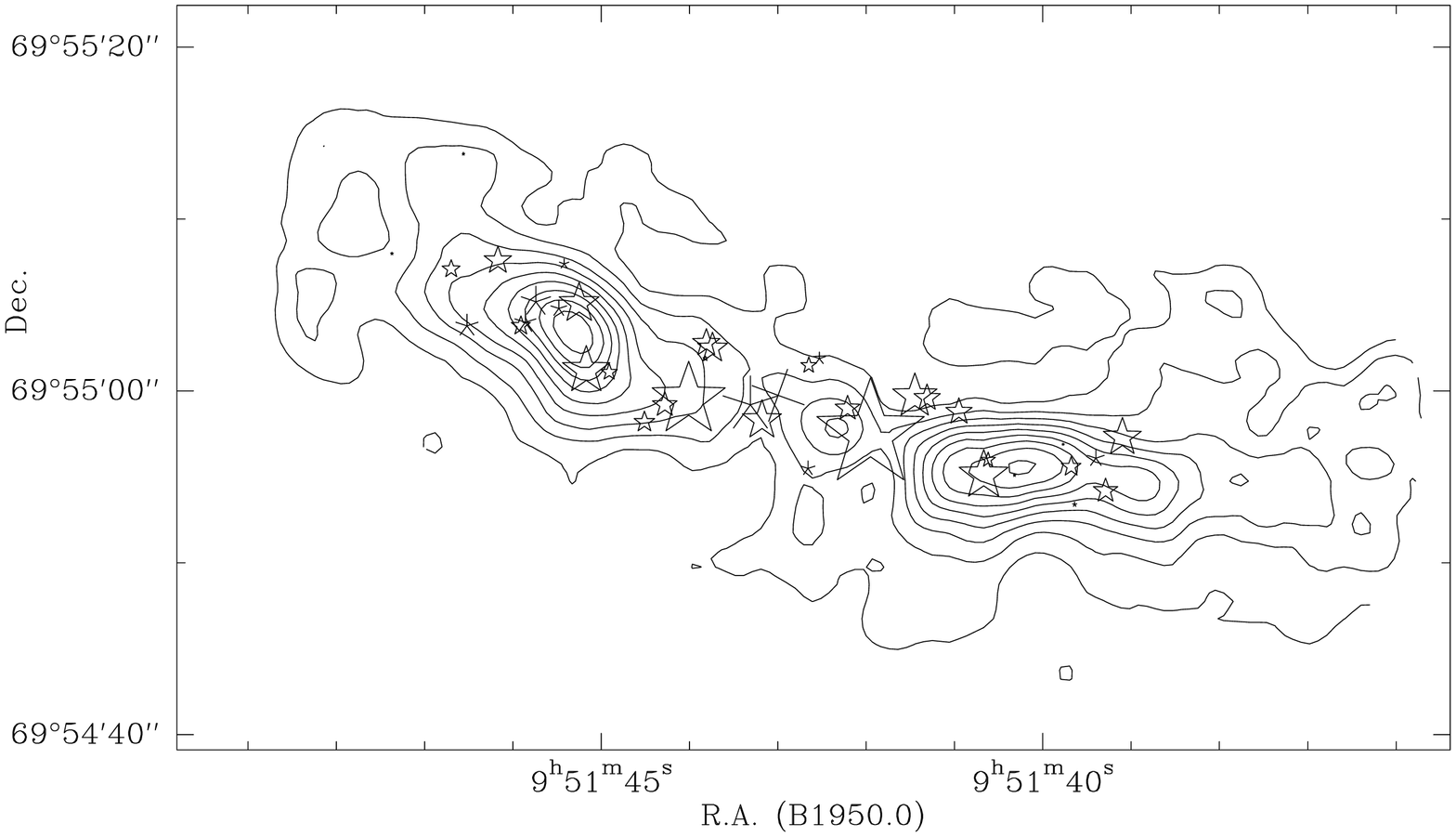,width=12cm,angle=270}\vspace{-4.7cm}}
\hfill\parbox[b]{5.5cm}{
\caption
{Map of the integrated $^{13}$CO(1$\rightarrow$0) emission of 
M\,82. Contour levels are 4, 6, 8... Jy/beam$\cdot$km\,s$^{-1}$. 
The cross marks the position of the $\lambda2~\mu$m nucleus
as in Fig.~2. The radio SNR found by Kronberg et al.\ (\cite{kronberg}) 
are indicated by stars whose sizes are proportional to the logarithm 
of the source flux densities; the SNR showing indications of 
free-free absorption (Wills et al.\ \cite{wills}) are denoted by
open stars.}
\label{fig:int13co}
}
\end{figure*}

Fig.~\ref{fig:int13co} shows the $^{13}$CO(1$\rightarrow$0) 
velocity-integrated line intensity (zeroth moment of the data cube).
Its distribution differs markedly from the continuum emission
distribution of Fig.~1. The two emission lobes appear much broader
and conspicuous than in the continuum emission, and the ``central''
source ($\alpha = 09^h51^m42\fs3$, $\delta = 69\degr54'57\farcs4$,
B1950.0), much fainter. Note that the map is corrected for attenuation by
the antenna primary beam and that its noise increases towards its edges.
The reality of the weak features near $\alpha = 09^h51^m36^s$, at the 
western edge, is thus questionable (see also Sect.~\ref{sec:velfi}). 

\begin{figure*}          
\vbox{   
\psfig{file=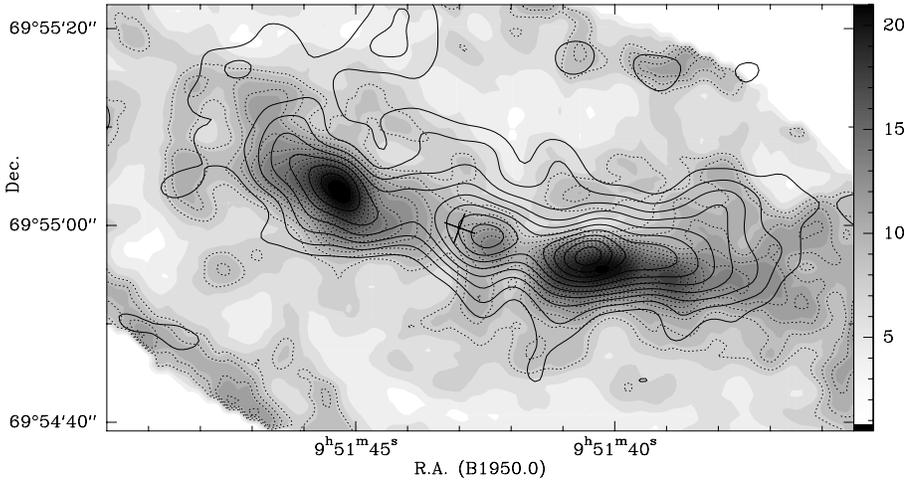,width=12cm,angle=270}\vspace{-4.5cm}}
\hfill\parbox[b]{5.5cm}{
\caption
{Shen \& Lo's $^{12}$CO velocity-integrated contours smoothed to 3\farcs5, 
superimposed on a grey-scale map of the $^{13}$CO integrated emission. 
First contour and contour interval are 25 Jy/beam$\cdot$km\,s$^{-1}$; 
the $^{13}$CO intensity code is indicated by the right scale and wedge 
(in Jy/beam$\cdot$km\,s$^{-1}$). Note that due to the relatively small 
size of the IRAM primary beam, the $^{13}$CO map becomes noisy 
at distances $> 30''$ from the major axis.}
\label{fig:isoratio} 
}
\end{figure*}

A comparison with the $^{12}$CO(1$\rightarrow$0) map of Shen \& Lo 
(\cite{shen}) shows good agreement of the most salient structures (see 
Fig.~\ref{fig:isoratio}). There is an almost perfect spatial coincidence 
between the bright lobes in $^{13}$CO and $^{12}$CO which holds as well
for the central peak, although this latter appears relatively brighter 
in $^{12}$CO than in $^{13}$CO. The faint outer extensions of the lobes 
in Figs.~\ref{fig:chann13} and \ref{fig:int13co} are also present in 
$^{12}$CO. The weak $^{13}$CO extension, south of the nucleus near 
$\alpha = 09^h51^m43\fs0$, $\delta = 69\degr54'52\farcs9$, has a 
counterpart in $^{12}$CO; this structure seems to be made up by emission 
at $\sim-50$~km\,s$^{-1}$ (see Fig.~\ref{fig:chann13}). Except for the 
stronger central source, the only marked difference between $^{12}$CO 
and $^{13}$CO  is the absence of $^{13}$CO emission to the N, at 
$\delta\sim 69\degr55'20''$. As a consequence, the  $R=^{12}$CO/$^{13}$CO 
(1$\rightarrow$0) line intensity ratio is about constant over most of the 
map and is close to 10, a value marginally larger than those observed in 
the bright nearby galaxies ($R= 7-10$, see e.g.\ Garc\'{\i}a-Burillo et al.\ 
\cite{santim51}). This ratio increases to $R=\simeq 20$ near the central CO source 
(C\,1 in Shen \& Lo's map), and could be even larger in the northern 
$^{12}$CO source. The minor differences between the $^{13}$CO and $^{12}$CO 
maps seem hardly compatible with the large ones reported by Loiseau et 
al.\ (\cite{loiseau9}). It has to be noted, however, that their analysis is based on
(2$\rightarrow$1) data so that a direct comparison is not possible. 

Comparing our $^{13}$CO(1$\rightarrow$0) with the HCN(1$\rightarrow$0) 
map of Brouillet \& Schilke (\cite{brouillet}) we find an overall 
resemblance of the large-scale distribution, but also significant 
differences.  Mainly, the central source is much more pronounced in
all velocity channels in HCN than in $^{13}$CO; conversely the lobes
appear dimmer. The faint 50~kms$^{-1}$ extension south of the nucleus
and the arc-like structure $7''$ west of the nucleus, discussed in the
next section, are also discernable, though less clearly. 
 
The $^{13}$CO data cast doubts on the standard picture of a circumnuclear
molecular torus. Firstly, the emission in Fig.~\ref{fig:int13co} is 
strongly asymmetric with respect to the major and minor axes.
Secondly, an edge-on torus should appear quite differently in the optically 
thin $^{13}$CO line and in the optically thick $^{12}$CO line. The 
$^{13}$CO(1$\rightarrow$0) line, for which $\tau \sim 0.1$ (Wild et al.\ 
\cite{wild}), should exhibit limb-brightening, i.e.\ two distant lobes, 
whereas the optically thick $^{12}$CO line ($\tau \sim 7$) should show a 
more uniform distribution, and emphasize the near side of the torus.  
Instead, we observe remarkably similar distributions, which suggests that
the CO is mostly concentrated in two or three sources. 

The observations seem better interpreted in terms of molecular gas 
condensations located at the ends and in the middle of a stellar bar. 
Such condensations or ``armlets'' have already been observed in several 
galaxies, including \object{NGC\,1530} (Reynaud \& Downes \cite{reynaud}) 
and \object{NGC\,891} (Garc\'{\i}a-Burillo \& Gu\'elin \cite{santimichel}); 
Achtermann \& Lacy (\cite{achtermann}) and Larkin et al.\ (\cite{larkin}) 
have already proposed a similar model in order to explain the optical 
forbidden lines at the centre of M\,82.  We will see in the next sections 
that the kinematical data support this interpretation.

\subsection{Kinematics of the $^{13}$CO gas} \label{sec:velfi}

\begin{figure*}     
\psfig{file=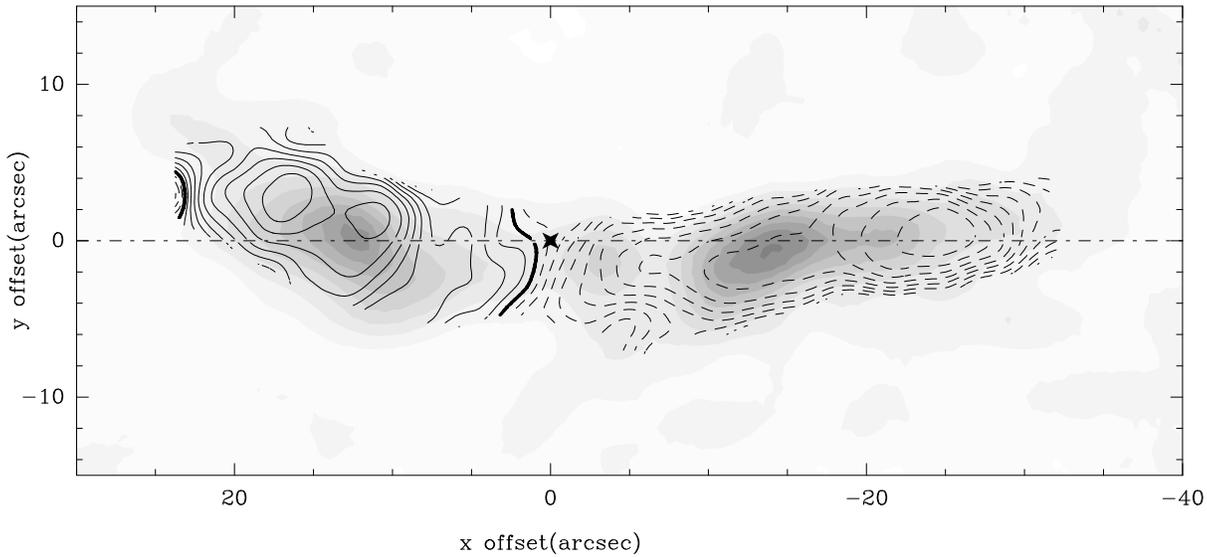,width=16cm,angle=270}
\caption
{Isovelocity contour levels, obtained from the first moment of 
$^{13}$CO(1$\rightarrow$0), overlaid on the integrated intensity map of 
Fig.~3 (grey scale). The thick line denotes $v=0$ (or $V_{sys}(LSR)= 
+225$~km\,s$^{-1}$) and the level step is 10\,kms$^{-1}$. Dashed contours 
indicate blue-shifted gas and full line contours red-shifted gas. $x$ 
represents the offset along the galaxy major axis (point-dashed line, 
{\it PA} $ = 75\degr$) and $y$ the offsets along the minor axis. $x$ is 
positive towards the E-NE and $y$ towards the the N-NW. The dynamical 
centre is indicated by a filled cross.}
\label{fig:velfi}
\end{figure*}

Fig.~\ref{fig:velfi} shows the velocity field derived from 
$^{13}$CO(1$\rightarrow$0) (first order moment), superimposed on the 
integrated intensity map.  The dominant rotation pattern shows the eastern 
side receding (red-shifted) and the western side approaching (blue-shifted). 
There is a strong velocity gradient around the dynamical centre 
(offset (0,0) in Fig.~\ref{fig:velfi}), and the iso-velocity contours are 
systematically tilted by $\sim 30\degr$ with respect to the minor axis.  
This tilt is visible along the whole major axis of M\,82, suggesting that 
the gas flow is driven by a non-axisymmetrical potential within the inner 
500\,pc. On the other hand, the velocity field is rather asymmetric, 
possibly due to the tidal interaction with \object{M\,81}. A detailed
analysis of individual features is thus difficult -- even the inclination
of the disk to the line of sight is hard to determine; analyses of the
outer disk and the outflow cones suggest a value of $i\sim 80\degr$ 
(G\"otz et al.\ \cite{mgoetz}).
  
The position angle ({\it PA}) of the major axis is slightly varying 
depending on the tracer and method used. That of the optical disk is 
usually given as $65\degr$, but G\"otz et al.\ (\cite{mgoetz}) used 
$60\degr$ for their thorough kinematical analysis of the optical emission 
lines. Other angles are also used, e.g.\ Achtermann \& Lacy 
(\cite{achtermann}) use $70\degr$ in the analysis of their Ne{\sc ii} data. 
We re-determined the position of the kinematical major axis of the molecular 
gas from our data cube together with the position of the dynamical centre 
(see Fig.~\ref{fig:velfi}). The {\it PA} of this axis is 
$75\degr$, $10\degr$ different from that of the outer disk. The dynamical
centre is found to coincide with the $\ \lambda 2\,\mu$m nucleus ($\alpha = 
09^h51^m43\fs4$, $\delta = 69\degr55'00\farcs0$ B1950.0,
 Joy et al.\ \cite{joy}).

\begin{figure*}      
\vbox{  
\psfig{file=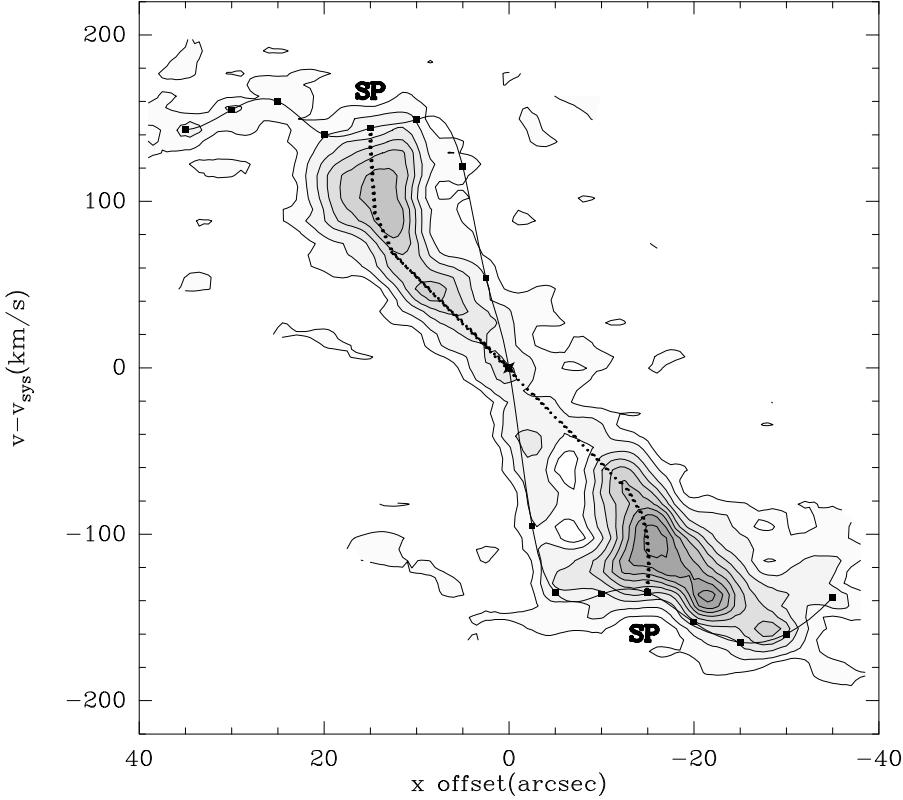,width=12cm,angle=270}\vspace{-4.1cm}}
\hfill\parbox[b]{5.5cm}{
\caption
{Position-velocity diagramme taken along the major axis (grey scale and 
line contours from 10 to 300\,mJy\,beam$^{-1}$ by steps of 
35\,mJy\,beam$^{-1}$). The S-shaped pattern denoted as {\bf SP} (dotted
curve) locates the signature of the gaseous spiral arm response to the 
stellar bar. The terminal velocities along the major axis (square markers 
connected by a line) serve to derive v$_{rot}$.}
\label{fig:posvel}
}
\end{figure*}

Fig.~\ref{fig:posvel} shows the position-velocity (p-v) diagramme taken 
along the major axis ($PA=75\degr$). Emission can be traced from 
$-190$~km\,s$^{-1}$ to $+130$~km\,s$^{-1}$. This implies a 
rotation speed -- {\it if this reflects simple rotation} -- of 
$165/\sin(i)$~km\,s$^{-1}$, exactly the same figure as derived by Shen 
\& Lo (\cite{shen}). We would then obtain the same indicative mass for 
the central $\sim800$~pc region of \object{M\,82} as Shen \& Lo 
($7\times10^8$ M$_{\odot}$). 

\begin{figure}          
\psfig{file=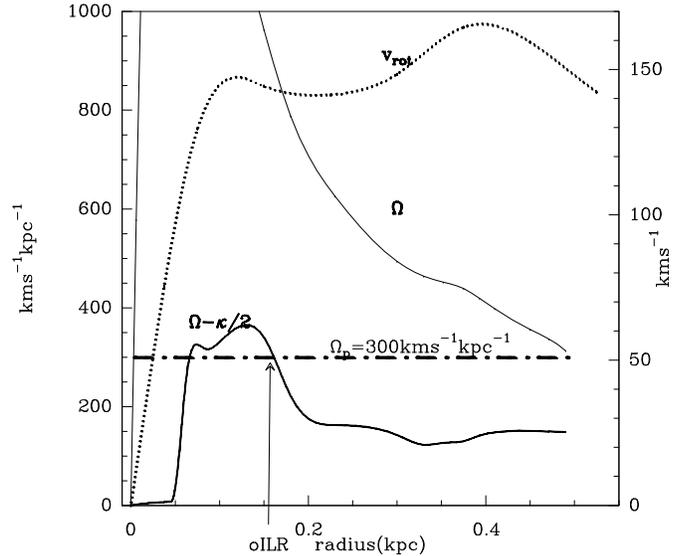,width=8.8cm,angle=270}
\caption
{Plot of the rotational velocity $v_{rot}$, of the angular frequency 
$\Omega$, and of $\Omega-\kappa/2$, the angular minus the epicyclic 
frequency, for the inner 500\,pc of M~82. For the assumed bar 
pattern speed of $\Omega_p=300$\,kms$^{-1}$\,kpc$^{-1}$, the oILR is 
at $r\sim150$\,pc.}
\label{fig:omega}
\end{figure}

Although the $^{13}$CO brightness distribution is very asymmetric, the main 
pattern of the p-v diagramme and the ridge of terminal frequencies are 
rather symmetric with respect to the dynamical centre. We note that 
emission from gas at high velocities is detected on both sides of the 
nucleus, although with a higher intensity for $x<0$. The rotation curve 
derived from the terminal-velocities method applied to p-v diagramme is very 
steep: we estimate $v_{rot}\sim 140$\,km\,s$^{-1}$ at $r\sim75$\,pc. From 
$r=75$\,pc to $r=300$\,pc the estimated rotation curve stays flat 
($v_{rot}=140$\,km\,s$^{-1}$) and reaches a relative maximum of 
$v_{rot}=160$\,km\,s$^{-1}$ at $r=375$\,pc. The rotation curve ($v_{rot}$) 
and the related principal frequencies derived in the epicyclic 
approximation ($\Omega$ and $\Omega$-$\kappa$/2) 
are represented in Fig.~\ref{fig:omega} for the inner 500\,pc. 

A large amount of gas is detected at velocities lower than 
determined by the ridge of terminal velocities shown in 
Fig.~\ref{fig:posvel}.  There is an S-shaped feature in the p-v 
diagramme going across the dynamical centre (denoted as SP in 
Fig.~\ref{fig:posvel}).  The characteristic pattern of SP 
suggests the presence of a non-axisymmetric distribution of molecular 
gas in the form of nuclear mini-spiral arms, which extend from the 
centre to $r\sim 300$\,pc.  Moreover, the `figure-eight' pattern of 
the p-v diagramme formed by SP and the curve of terminal velocities is 
typical of a bar-driven gas flow (Kuijken \& Merrifield  
\cite{kuijken}).  The existence of a nuclear stellar bar of 1\,kpc 
diameter was already established by Telesco \& Gezari 
(\cite{telesco9}), based on near-infrared observations in the J, K and 
I bands.  Driven by a bar potential, gas clouds follow non 
self-intersecting ellipsoidal orbits.  The major axes of these orbits 
precess as a function of radius, and hence end up delineating spiral 
arms, owing to orbit crowding.  The precession of gas orbits is due to 
the dissipative nature of gas: molecular cloud-cloud collisions and 
the implied viscosity of the process cause a smooth transition between 
the bar-driven $x_1$ orbits (parallel to the bar major axis) towards 
$x_2$ orbits (perpendicular to bar major axis), when we go across the 
Inner Lindblad Resonance (ILR).  Therefore, the presence of a nuclear 
bar potential and a spiral gas response are intimately related.

The existence of two ILRs in the nucleus of \object{M\,82} is clearly 
suggested by our observations (see Fig.~\ref{fig:omega}). The ring-like 
appearance of the integrated intensity map (referred to in the literature 
as the molecular gas torus) indicates gas accumulation towards the outer 
ILR (oILR) at $r\sim150$\,pc caused by the action of gravitational torques 
by the stellar bar on the gas. If we assume that corotation of the 
stellar bar is located near its end-points ($r\sim500$\,pc), we derive 
a bar pattern speed of $\Omega_p=300$\,kms$^{-1}$\,kpc$^{-1}$ and from 
it, the location of two ILRs at $r$(iILR)\,$\sim50\ldots100$\,pc and 
$r$(oILR)$\sim150\ldots200$\,pc (see Fig.~\ref{fig:omega}). The onset 
of a fast bar instability in the nucleus of \object{M\,82} is hardly 
surprising in view of the measured high values of $\Omega-\kappa/2$ 
within the inner 500\,pc. Results of recent near-infrared surveys of 
galactic centres show that nuclear bar instabilities are ubiquitous 
and that the pattern speeds of these m=2 instabilities can reach high 
values on the derived kinematical major axis (see the case of 
\object{M\,100}: Garc\'{\i}a-Burillo et al.\ \cite{santimaria}). The 
final output would be the formation of a two-arm gaseous trailing spiral 
across the oILR. The tilting of isovelocity contours towards the centre, 
the ring-like concentration of molecular gas and the `figure-eight' 
pattern of the major axis p-v plot support this theoretical scenario 
for the nucleus of \object{M\,82}.

The steep negative-velocity feature between $x=0$ and $x= -5''$, which
is linked to the bar, was also observed in HCN(1$\rightarrow$0) by 
Brouillet \& Schilke (\cite{brouillet}) and in $^{12}$CO 
(1$\rightarrow$0) by Shen \& Lo (\cite{shen}). It is more pronounced 
in HCN than in $^{12}$CO or $^{13}$CO, which probably denotes that the 
gas near the centre of the bar is dense. 

\section{A giant bubble} \label{sec:arc}

This steep negative-velocity feature merges into the western lobe at 
$-12''$, leaving a prominent hole in the p-v diagramme at $x=-7''$, 
${\rm v= -50\  to -100\, kms^{-1}}$. This ``hole'' is also visible in
the velocity-channel maps of Fig.~\ref{fig:chann13}, where it
appears as a $5''$-wide gap between the western lobe and the central 
source and centred at $\Delta \alpha = 7'', \Delta \delta = -2''$.  

\begin{figure*}       
\vbox{ 
\psfig{file=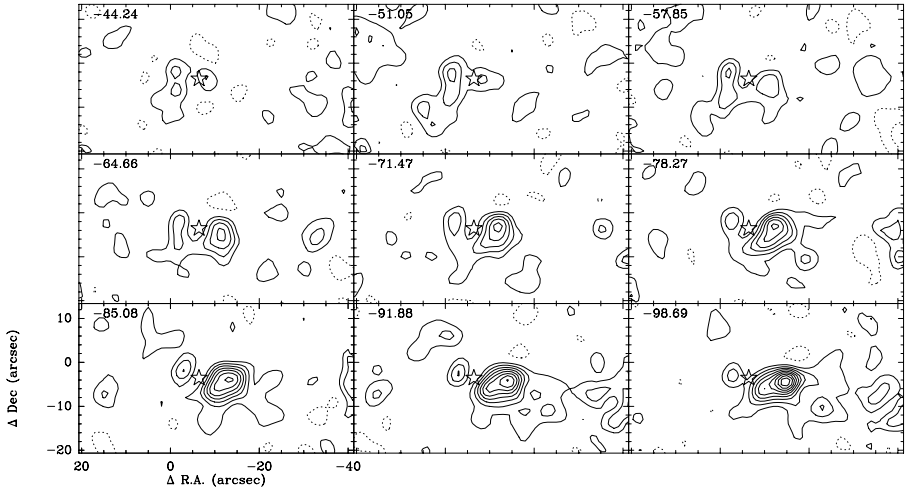,width=12cm,angle=270}\vspace{-2.9cm}}
\hfill\parbox[b]{5.5cm}{
\caption
{Channel maps between $-44$~km\,s$^{-1}$ and $-99$~km\,s$^{-1}$, 
centred on the emission hole and the arc-like feature. The star
markes the position of the brightest compact continuum source in M\,82, 
41.9+58, located at $\alpha = 09^h51^m42\fs0$, $\delta = 69\degr54'57^s4$.}
\label{fig:arcchann} 
}
\end{figure*}

Fig.~\ref{fig:arcchann} shows a close-up view of velocity-channel maps 
around this hole. A faint arc-like structure (diameter $\sim 8''$, 
or 130~pc) delineates the southern border of the hole, while only very
weak emission is visible to the north (at $-51$ km\,s$^{-1}$).  
The hole happens to host the young SNR 41.9+58 which is the strongest 
compact radio source in M~82 (Kron\-berg et al.\ \cite{kronberg}). This 
SNR is however assumed to be only about 40 years old (Wilkinson \&  
de Bruyn \cite{wilkinson}) and hence cannot be the origin of such a 
large structure. We note that the arc-like structure and the hole are
also visible in the $^{12}$CO velocity-channel maps shown 
by Lo et al.\ (\cite{lo}) (e.g.\ in their map at $V_{{\rm LSR}} = 
172$~km\,s$^{-1}$, which corresponds to $v= -53$~km\,s$^{-1}$). 

The gas in the CO hole seems largely ionized. The H41$\alpha$ (Seaquist 
et al.\ \cite{seaquist8}) and H$110\alpha$ (Seaquist et al.\ 
\cite{seaquist9}) recombination line emissions show both a maximum,  
and the $4'' \times 7''$ wide associated H{\sc ii} region fills the hole 
in our $^{13}$CO channel maps. The 3~mm continuum emission, which is 
essentially thermal free-free radiation, also peaks at this position. As 
concerns molecular lines, Wild et al.\ (\cite{wild}) reported emission 
of high-J lines of HCO$^+$ and HCN  from the region of the arc+hole, 
while lower excitation lines are less pronounced. The maximum of the 
atomic C{\sc i} line falls close to this location (White et al.\ 
\cite{white}). Those authors also argued that the enhanced C{\sc i} 
abundance in the central region of \object{M\,82} can be connected to 
a higher cosmic ray flux there. Copious cosmic rays could be delivered 
e.g.\ by the luminous compact source 41.9+58. Furthermore, as noted by 
Wild et al.\ (\cite{wild}), the secondary peak in the 2.2$\mu$m image  
as well as the FIR and submm dust emission peaks (Joy et al.\ \cite{joy}; 
Smith et al.\ \cite{smith}) are found right there, and the maxima of the 
[Ne{\sc ii}] line (Achtermann \& Lacy \cite{achtermann}) are both close 
to this position. Finally, a recent high-resolution map at 408\,MHz 
(Wills et al.\ \cite{wills}) clearly shows an emission-free circle of 
100\,pc diameter, right around this source, which is ascribed to a large 
photoionized zone. A realistic interpretation is that this is a region 
with an intense radiation field in which the atomic hydrogen has been 
ionized (see Wills et al.\ \cite{wills}). Such a region might have been
created by a large cluster of massive stars whose stellar winds and 
supernova explosions (of which 41.9+58 is just the latest) can easily 
clear a large hole in the interstellar matter. Therein hot gas will be 
the dominant constituent and the intense radiation of the remaining stars 
will dissociate molecules unless they are shielded within dense clouds. 
This is consistent with the finding that the gas deficiency is less 
pronounced here in the HCN line (Sect.~\ref{sec:line13co}). 

\section{The fueling of the starburst} \label{sec:discuss}

In order to try to understand the remarkable morphology and kinematics of the 
central 700~pc region we briefly describe what could have happened since 
the probable close encounter of \object{M\,82} with \object{M\,81}. Cottrell 
(\cite{cottrell}) interpreted the observed large-scale kinematics of 
the neutral hydrogen gas in terms of tidal disruption of this gas from 
\object{M\,81} during the passage of \object{M\,82} on a hyperbolic 
orbit. The fact that the kinematic axis of the H{\sc i} surrounding 
\object{M\,82} lies parallel to the major axis was seen as evidence 
for the captured gas to be in a polar orbit around \object{M\,82}. In 
this picture, gas would eventually fall into the centre of \object{M\,82} 
and thus feed the starburst of this presumably former gas-poor galaxy. 
 
Recently, Yun et al.\ (\cite{yun}) have proposed the opposite scenario: the 
large H{\sc i} streamers which they have found are interpreted as gas 
torn out of the outer gas-rich H{\sc i} disk of \object{M\,82}. The question 
arises how in this case gas is transported to the centre to feed the 
starburst. There is of course no contradiction because the tidal 
forces that are responsible for gas disruption in the outer disk of 
\object{M\,82} will also have caused instabilities in the gas orbiting in its 
inner part. It is well known that tidal interactions induce the rapid 
formation of bars along which gas can then be transported to the central 
regions of the galaxies. In view of this, the hypothesis of Yun et al.\ 
seems more attractive, because in this scenario gas may be transported 
more efficiently to the centre of \object{M\,82} than from a polar orbit. 
We therefore investigate in the following whether the observed kinematics 
of the $^{13}$CO emission is consistent with this view. 

The maps tracing the molecular gas in \object{M\,82} have been mostly
interpreted in terms of a rotating molecular torus (e.g.\ Nakai et 
al.\ \cite{nakai}; Loiseau et al.\ \cite{loiseau8}; Shen \& Lo \cite{shen}).
As pointed out in Sect.~\ref{sec:line13co}, the $^{13}$CO high resolution
maps reveal a patchy gas distribution and a disturbed kinematics 
characteristic of gas in orbit around a stellar bar. The bar forces the outer 
gas to flow towards the central region, where it is exposed to the intense 
radiation field produced by the starburst and becomes largely dissociated.
This is when the denser and better shielded clouds get shaped into the 
two lobes and the central condensation of Fig.~3.

\begin{figure}     
\vspace*{6mm}
\hspace*{28mm}
\vbox{  
\psfig{file=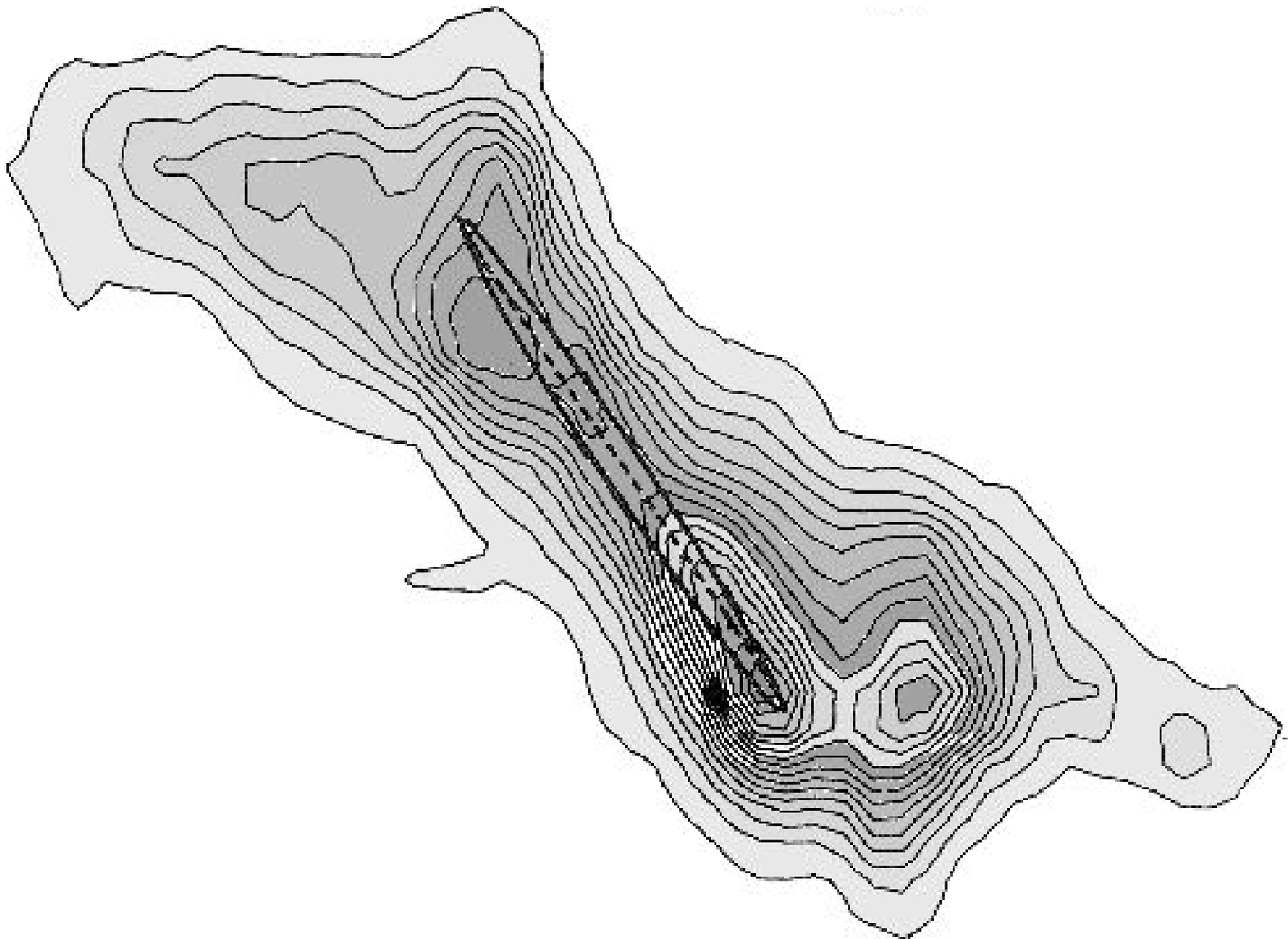,width=42.2mm,height=73.2mm,angle=90}
\vspace*{-78mm}\hspace*{-28.9mm}
\psfig{file=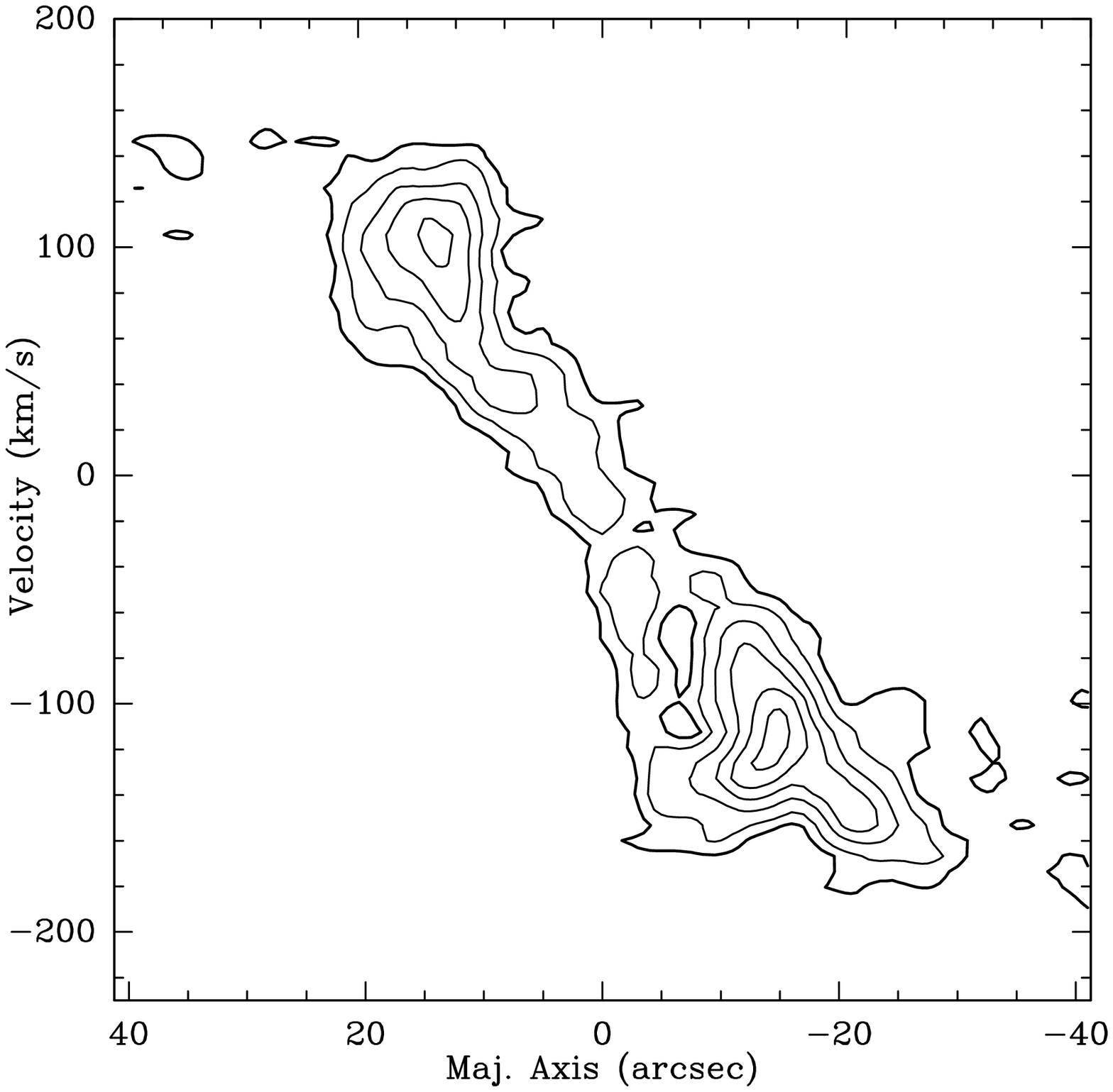,width=8.8cm,angle=0}
}
\caption
{Position-velocity diagramme of the [Ne{\sc ii}] emission 
(adapted from Achtermann \& Lacy \cite{achtermann} -- dense, shaded 
contours), superimposed onto that of the $^{13}$CO(1$\rightarrow$0) 
emission (thicker, widely spaced contours), and centred on 
$\alpha = 09^h51^m43\fs4$, $\delta = 69\degr55'00\farcs0$, with 
position angle +70$\degr$. The [Ne{\sc ii}] `ring' as traced by the 
narrow ellipse has its endpoints in regions void of $^{13}$CO emission.} 
\label{fig:pvcomp}
\end{figure}

We have compared several p-v diagrammes and find that they support 
this picture.  These are: $^{13}$CO (Fig.~\ref{fig:posvel}), [Ne{\sc 
ii}] (Achtermann \& Lacy \cite{achtermann}; their Fig.~4), 
H166$\alpha$ (Roelfsema \& Goss \cite{roelfsema}), H41$\alpha$ 
(Seaquist et al.\ \cite{seaquist9}; their Fig.~5c), HCO$^+$ (Seaquist 
et al.\ \cite{seaquist9}; their Fig.~5), and HCN (Brouillet \&  
Schilke \cite{brouillet}; their Fig.~5; Shen \& Lo \cite{shen}; their 
Fig.~4).  From this we summarize the following: The brightest [Ne{\sc 
ii}] emission peaks (concentrated between $\Delta \alpha +6''$ 
and $-11''$ around $-110$\,km\,s$^{-1}$, see Fig.~\ref{fig:pvcomp}) 
coincide with depressions in $^{13}$CO. The same holds true for the 
radio recombination lines, which shows that the effect does not just
result from visual extinction variations. Finally, HCO$^+$(1$\rightarrow$0)
closely follows the $^{13}$CO emission and HCN(1$\rightarrow$0) is 
brighter near the nucleus and shows little evidence of a hole 
$7''$ W of it.

The velocity-integrated emission maps of these lines show that the 
tracers of current star formation and the ionized gas are confined to 
radii of $\le10'' \ldots 14''$, which is also seen in the 3-mm continuum 
(free-free radiation) and the mid-infrared, the latter reflecting the 
heating of dust by young stars (Telesco \& Gezari \cite{telesco9}).  
In contrast, the bulk of the CO gas is located beyond $\sim10''$.  
The gas traced by the HCN molecule resides in very dense clouds 
that are better shielded against the radiation field.

\section{Summary and conclusions} \label{sec:sum}

We have performed high-resolution observations of \object{M\,82} 
in the $^{13}$CO(1$\rightarrow$0) line using the Plateau de Bure 
interferometer. The distribution and kinematics of this molecular 
species have been derived and analyzed, and a first comparison is made 
with existing interferometric data in the lines of 
$^{12}$CO(1$\rightarrow$0) and HCN(1$\rightarrow$0).

The complex kinematic structure unveiled in previous studies is 
confirmed.  Together with the asymmetric morphology of the observed 
distribution of the molecular gas it makes it difficult to maintain the 
hypothesis of an edge-on molecular torus, an idea advanced 
when the first low-resolution single-dish maps of the 
CO gas in the centre of \object{M\,82} became available.  We rather 
propose that what we observe is the signature of a bar, 
the central (projected) 200~pc portion of which is relatively void of 
CO gas and probably subject to strong dissociation.

A 130 pc-wide emission hole is seen in our $^{13}$CO data cube; it
coincides with a region of enhanced high-J lines, recombination 
lines emission peak, and strong free-free and C{\sc i} 
emission. It also hosts the young SNR 41.9+58.  We think that 
this CO hole reflects a bubble inside which the gas is 
ionized and the molecules dissociated. 

Maps of several transitions of the rare CO isotopomers would be 
needed to better constrain the cloud properties. Together with
one arcsec resolution maps of CO and HCN, which could be directly 
compared to the optical pictures, they could yield a
real understanding of how is triggered and fueld the most 
spectacular starburst in the vicinity of the Galaxy.

\acknowledgements It is a pleasure to thank Drs.\ J. Shen and K.Y. Lo 
for making available to us their $^{12}$CO(1-0) data from BIMA. We are 
very grateful to the referee, Dr.\ D. Jaffe, for his many helpful 
suggestions. UK is very indebted to IRAM for the warm hospitality and 
financial support.

\end{document}